\documentclass[lettersize,journal]{IEEEtran}
\usepackage{amsmath,amssymb,amsfonts}
\usepackage[linesnumbered,ruled,vlined]{algorithm2e}
\usepackage{array}
\usepackage{textcomp}
\usepackage{stfloats}
\usepackage{url}
\usepackage{verbatim}
\usepackage{graphicx}
\usepackage{cite}
\usepackage{multirow}
\usepackage{multicol}
\usepackage{subfigure}
\usepackage{pgfplots}
\usepackage{tikz}
\usepackage{tkz-tab}
\usepackage{tabularx}
\usepackage{float}
\usetikzlibrary{patterns}
\usepgfplotslibrary{statistics}
\pgfplotsset{compat=1.18}
\pgfmathdeclarefunction{fpumod}{2}{%
    \pgfmathfloatdivide{#1}{#2}%
    \pgfmathfloatint{\pgfmathresult}%
    \pgfmathfloatmultiply{\pgfmathresult}{#2}%
    \pgfmathfloatsubtract{#1}{\pgfmathresult}%
    \pgfmathfloatifapproxequalrel{\pgfmathresult}{#2}{\def\pgfmathresult{5}}{}%
}
\pgfplotsset{boxplot legend/.style={
    legend image code/.code={
        \draw[#1] (0cm,-0.1cm) rectangle (0.4cm,0.1cm)
        (0.2cm,-0.1cm) -- (0.2cm,-0.2cm) (0.05cm,-0.2cm) -- (0.35cm,-0.2cm)
        (0.2cm,0.1cm) -- (0.2cm,0.2cm) (0.05cm,0.2cm) -- (0.35cm,0.2cm);
     \path (0cm,0.24cm) (0cm,-0.24cm);  
    },
}}

\begin{document}

\title{An Intelligent Quantum Cyber-Security Framework for Healthcare Data Management}

\author{
Kishu Gupta,~\IEEEmembership{Member,~IEEE,}  Deepika Saxena,~\IEEEmembership{Member,~IEEE,} \\Pooja Rani, Jitendra Kumar,~\IEEEmembership{Senior member,~IEEE,} \\Aaisha Makkar,~\IEEEmembership{Member,~IEEE,} Ashutosh Kumar Singh,~\IEEEmembership{Senior member,~IEEE,} \\and Chung-Nan Lee,~\IEEEmembership{Member,~IEEE}
\thanks{Manuscript received 20 June 2024; accepted 4 September 2024. This article was recommended for publication by Associate Editor M.-H. Hung and Editor X. Xie upon evaluation of the reviewers’ comments. This work was supported by the National Sun Yat-sen University, Kaohsiung, Taiwan. (Corresponding author: Deepika Saxena.)}
\thanks{Kishu Gupta and Chung-Nan Lee are with the Department of Computer Science and Engineering, National  Sun Yat-sen University, Kaohsiung, 80424, Taiwan (e-mail: kishuguptares@gmail.com, cnlee@mail.cse.nsysu.edu.tw).}
\thanks{Deepika Saxena is with the School of Computer Science and Engineering, The University of Aizu, Aizuwakamatsu 965-0006, Japan (e-mail: deepika@u-aizu.ac.jp).}
\thanks{Pooja Rani is with the Department of Computer Applications, National Institute of Technology Kurukshetra, Kurukshetra 136119, India (e-mail: poojavats971993@gmail.com).}
\thanks{Jitendra Kumar is with the Department of Mathematics, Bioinformatics, and Computer Applications, Maulana Azad National Institute of Technology, Bhopal, Bhopal 462003, India (e-mail: jitendrakumar@ieee.org).}
\thanks{Aaisha Makkar is with the College of Computer Science and Engineering, University of Derby, DE22 3AW Derby, U.K. (e-mail: a.makkar@derby.ac.uk).}
\thanks{Ashutosh Kumar Singh is with Department of Computer Science and Engineering, Indian Institute of Information Technology Bhopal, Bhopal 462003, India, and also with Department of Computer Science, the University of Economics and Human Sciences, 01-043 Warsaw, Poland. (e-mail: ashutosh@iiitbhopal.ac.in).}
\thanks{Color versions of one or more figures in this article are available at https://doi.org/10.1109/TASE.2024.3456209.
\\Digital Object Identifier 10.1109/TASE.2024.3456209}}


 \markboth{IEEE TRANSACTIONS ON AUTOMATION SCIENCE AND ENGINEERING}%
{Shell \MakeLowercase{\textit{Gupta et al.}}: A Sample Article Using IEEEtran.cls for IEEE Journals}

\makeatletter
\newcommand{\removelatexerror}{\let\@latex@error\@gobble}
\def\ps@IEEEtitlepagestyle{%
	\def\@oddfoot{\mycopyrightnotice}%
	\def\@oddhead{\hbox{}\@IEEEheaderstyle\leftmark\hfil\thepage}\relax
	\def\@evenhead{\@IEEEheaderstyle\thepage\hfil\leftmark\hbox{}}\relax
	\def\@evenfoot{}%
}

\def\mycopyrightnotice{%
	\begin{minipage}{\textwidth}
		\centering \scriptsize
		1545-5955©2024 IEEE. Personal use is permitted, but republication/redistribution requires IEEE permission.
		See https://www.ieee.org/publications/rights/index.html for more information.\\This article has been accepted in IEEE Transactions on Automation Science and Engineering Journal © 2024 IEEE. Personal use of this material is permitted. Permission from IEEE must be obtained for all other uses, in any current or future media, including reprinting/republishing this material for advertising or promotional purposes, creating new collective works, for resale or redistribution to servers or lists, or reuse of any copyrighted component of this work in other works. This work is freely available for survey and citation.
		
	\end{minipage}
}
\makeatother

\maketitle

\begin{abstract} 
Digital healthcare is essential to facilitate consumers to access and disseminate their medical data easily for enhanced medical care services. However, the significant concern with digitalization across healthcare systems necessitates for a prompt, productive, and secure storage facility along with a vigorous communication strategy, to stimulate sensitive digital healthcare data sharing and proactive estimation of malicious entities. In this context, this paper introduces a comprehensive quantum-based framework to overwhelm the potential security and privacy issues for secure healthcare data management. It equips quantum encryption for the secured storage and dispersal of healthcare data over the shared cloud platform by employing quantum encryption. Also, the framework furnishes a quantum feed-forward neural network unit to examine the intention behind the data request before granting access, for proactive estimation of potential data breach. In this way, the proposed framework delivers overall healthcare data management by coupling the advanced and more competent quantum approach with machine learning to safeguard the data storage, access, and prediction of malicious entities in an automated manner. Thus, the proposed IQ-HDM leads to more cooperative and effective healthcare delivery and empowers individuals with adequate custody of their health data. The experimental evaluation and comparison of the proposed IQ-HDM framework with state-of-the-art methods outline a considerable improvement up to 67.6\%, in tackling cyber threats related to healthcare data security. \\

\textbf{\textit{Note to Practitioner--}} \textbf{This paper aims to address the issue of digital healthcare data access, which requires both ease and security. Existing research either focuses solely on safe access or on high security, which often comes with high computational challenges. In this paper, we present a comprehensive approach that takes into account various challenges such as secure data storage, efficient data communication, and the prediction of malicious entities. We have developed a mathematical system to portray the overall management of healthcare data. All techniques proposed in this paper have been implemented using quantum computing and have been tested on four healthcare datasets. Initial experimental results suggest that the proposed approach is feasible. Our techniques can be applied to discover malicious entities and understand the behavior of real-life users in healthcare processes.
}
\end{abstract}


\section{Introduction}\label{secint}
\IEEEPARstart{D}{igital} transformation of healthcare services allows access to healthcare facilities, across the globe even from far distant, remote, and strategic locations. Cloud computation serves as a base element to roll out this worldwide digital facility by offering outstanding services like storage, computation, investigations, analysis, etc. at a very nominal cost and with remarkable availability. This has attracted healthcare institutions to migrate their data accumulated from diverse medical IoT devices and sensors, over the cloud platform \cite{TASE1-9237105, FedMUPGUPTA2024111519, gupta_Kush_2020-infocomp}. Cloud service platforms act as the backbone of digital health systems by enabling the digital retrieval of patient data and the extraction of valuable clinical information. As a result, various additional uses have become available, including quality management, healthcare administration, and trans-national research \cite{TC9471010, deps2-10272307}. The adoption of digital infrastructure by healthcare organizations can offer numerous advantages for consumers like doctors, patients, and healthcare services. However, apprehensions regarding the privacy and security of end consumer data is a major challenge across various healthcare institutions \cite{Pan10254468, ISoga9813496}. These institutions are required to grant data access among multiple stakeholders, such as researchers, academia, doctors, patients, regulatory bodies, etc., for different usages \cite{TC8870212, GDSPS10461067, MAIDS, TASE2-9966651}. Sharing of this crucial and sensitive data in digital infrastructure is essential for medical growth, but it is highly susceptible to data breaches, security, and privacy issues. According to a survey, healthcare data breaches have consistently trended upward and doubled in the last three years \cite{HDS}. Moreover, according to the Global Threat Report 2023, for more than 2500 adversaries there is a 112\% increase in the cyber threat-related eCrimes, compared to 2021 \cite{GTR2023}. Furthermore, a study highlights the privacy security concerns related to Electronic Health Records (EHRs) \cite{c6mishra}. Any mal-intentional exposure of crucial medical data to some unauthorized party may induce direct financial loss, reputation damage, operational downtime, legal actions and massive harm to overall growth of the organization's \cite{ICL8010262, TC9003529}. Thus, data breach outcomes are wide-ranging and extremely impactful. In this way, fortified data storage, reliable communication, edge security, and privacy appear as crucial challenges in shared cloud environments that must be handled properly. Proactive, healthcare data breach estimation emerges as a prominent way out of this problem. Several approaches \cite{5487521, gupta2020mlpam, IOTHSM9439861,TASE3-9474579,TASE4-9758074} have been defined in this regard but these approaches detect data breaches after their occurrence. However, in a real environment, proactive computation of possible data breaches is the key to safeguarding healthcare confidential data security and privacy.

In this context, a novel \textbf{I}ntelligent \textbf{Q}uantum Cyber-Security Framework for \textbf{H}ealthcare \textbf{D}ata \textbf{M}anagement (\textbf{IQ-HDM}) framework is proposed to accomplish the comprehensive healthcare data management by equipping secure storage and communication to estimate the crucial healthcare data access intention for being `malicious' or `non-malicious' and identification of malicious entity, in case of data breach. To the best of author's knowledge, this is the first framework that concurrently addresses the aforementioned multiple data security issues by furnishing a \textit{Quantum one-time padding encryption} (QOTPE) unit for secure data storage and \textit{Quantum feed-forward neural network} (QFNN) based \textit{quantum-protected healthcare data communication} (QPHDC) request analysis unit for proactive estimation of a mal-intentional entity. Thus, the proposed IQ-HDM framework establishes a comprehensive quantum-oriented security-embedded automated healthcare data management. 

\subsection{Related Work}\label{subsecrel}
 The considerable works presented for preserving healthcare data security via privacy-preserving, encryption, and prediction approaches, a few acting in reactive and others in a proactive manner. Lim et al. \cite{privatelink8742539} presented a more practical, scaleable, and easy-of-deploy solution to address the problem of privacy-preserving dataset integration using the concept of a prototype called PrivateLink without requiring key sharing among participants. Gupta et al. \cite{Forecast10.1007/978-981-15-8335-3_1} proposed a forecasting-based data leakage prevention (DLP) model to restrict data access permission to users by using a simple piece-wise linear function for model learning. This approach forecast possible guilty users based on past data access records of users. Rosa et al. \cite{TC9471010} presented a small-form and battery-less implantable device with acquisition channels for bio-potential, arterial pulse oximetry, and temperature recordings with in-situ encryption of data. Though implantable devices are the future of the remote medical field but they suffer from data theft and spoofing. Data disclosure poses serious threats to data security. This concept lags in strict data security norms. Gupta et al. \cite{gupta2020mlpam} proposed a novel model to support multiple participants to securely share their data for distinct purposes. The model defines the access policy and communication protocol among the involved multiple untrusted parties by utilizing encryption, machine learning, and probabilistic approaches. Xu et al. \cite{IOTHSM9439861} presented a model to process the complex healthcare security event in real-time by analyzing the security performance, using an improved convolutional neural network (CNN) having a four-branch inception block to increase the width of the CNN while reducing the parameters.
 
 Gupta et al. \cite{QM-MUP9865138} proposed a novel quantum machine learning based malicious user prediction (QM-MUP) and privacy-preserving model. The proposed model preserves data privacy via the Laplace mechanism-based noise addition and uses Quantum Pauli gate-based Neural Network predictor to exploit the computational and behavioral properties of qubits. Sun et al. \cite{PBAC-FG9647918} proposed privacy-preserving bilateral fine-grained access control (PBAC-FG) which employs fine-grained access control and matchmaking encryption technologies to ensure participants can specify their respective fine-grained access control over the encrypted healthcare data. Thus allowing only authorized counterparts to access the healthcare data. Song et al. \cite{COAB-PRE10098652} proposed a cryptographic approach, controllable out-sourced attribute-based proxy re-encryption (COAB-PRE) enabling bilateral and distributed access control whereby data producers and data consumers can both specify policies the other party must satisfy without a centralized access control server along with supporting verifiability to find out a wrong result produced by the edge nodes and locate the misbehaved one. Chang et al. \cite{DQFHE-CHANG2023103510} proposed a universal quantum circuit (UQC) based scheme named DQFHE to deal with the volatility problem of the servers using a quantum environment. Gupta et al. \cite{MAIDS} proposed a novel malicious user detection model by using the gradient boosting. Also, Gupta et al. \cite{FedMUPGUPTA2024111519} proposed a data security model to predict the malicious user in advance. It utilized the federated machine learning which incorporated data safety by making the learning at local user site without actually sharing data. Table \ref{tab:tabrelwork} showcases the studied literature in a consolidated form.
\begin{table}[!ht]
\caption{Encapsulated view of Related studies and proposed work}
 \label{tab:tabrelwork}
\centering
\resizebox{0.99\columnwidth}{!}{
\begin{tabular}{|c|c|c|c|c|l|}
\hline
\multirow{2}{*}{\textbf{Contributor}} & \multicolumn{3}{c|}{\textbf{Features}} & \multirow{2}{*}{\textbf{Strategy}} & \multirow{2}{*}{\textbf{Target}} \\ 
\cline{2-4} & $\mathbb{E}$ & $\mathbb{P}$ & $\mathbb{D}$ & &  \\
\hline
\hline
Lim et al. \cite{privatelink8742539} & $\checkmark$ & $\times$ & $\times$ & PP & Secure data sharing \\ \hline
Gupta et al. \cite{Forecast10.1007/978-981-15-8335-3_1} & $\times$ & $\checkmark$ & $\times$  & Forecasting & Malicious party detection \\ \hline
Gupta et al. \cite{gupta2020mlpam} & $\checkmark$ & $\times$ & $\checkmark$ & Probabilistic & Guilty user detection \\ \hline
Xu et al. \cite{IOTHSM9439861} & $\times$ & $\checkmark$ & $\times$ & I-CNN & Healthcare security \\ \hline
Gupta et al. \cite{QM-MUP9865138} & $\times$ & $\checkmark^q$ & $\times$ & QNN & Malicious user prediction \\ \hline
Sun et al.\cite{PBAC-FG9647918} & $\checkmark$ & $\times$ & $\times$ & PP & Unauthorized access control \\ \hline
Song et al. \cite{COAB-PRE10098652} & $\checkmark$ & $\times$ & $\times$ & ABE & Data access control \\ \hline
Chang et al. \cite{DQFHE-CHANG2023103510} & $\checkmark^q$ & $\times$ & $\times$ & HE, UQC & Server management issue \\ \hline
Gupta et al. \cite{MAIDS} & $\times$ & $\times$ &$\surd$ & XG-Boost & Malicious user detection  \\ \hline
Gupta et al. \cite{FedMUPGUPTA2024111519} & $\times$ & $\surd$ & $\times$ & Federated Learning & Malicious user prediction  \\ \hline
\textbf{IQ-HDM} & $\checkmark^q$ &$ \checkmark^q$ & $ \checkmark^q$ & QOTPE, QFNN & Healthcare data management \\ \hline \hline
\end{tabular}}

\footnotesize{$\checkmark^q$: Quantum mechanics, $\mathbb{E}$: Encryption, $\mathbb{P}$: Prediction, $\mathbb{D}$: Detection, PP: Privacy preserving, ABE: Attribute based encryption, HE: Homomorphic encryption, UQC: Universal quantum circuits, QNN: Quantum neural network, I-CNN: Improvised convolutional neural network, QE: Quantum encryption, QFNN: Quantum feed-forward neural network.}
\end{table}
\par The approaches discussed above assure data security, either using encryption to provide secure data access, prediction of malicious entities, or detection of guilty agents. Thus, existing work has presented data safety solutions for addressing a particular data security issue. Moreover, all these issues like secure data storage, data communication, and malicious entity prediction should be handled together, as these are diverse constituents of comprehensive data security. However, none of the existing work is sufficient to tackle the above-mentioned issues, concurrently considering these issues are different components of security. Unlike existing work, a comprehensive framework with a more potent, computational, and enhanced performance is proposed which addresses the limitation of the existing work to deliver data security. The proposed framework mitigates the malicious data request proactively to shield data from further breaches by utilizing the capability of QOTPE unit and QPHDC unit. Quantum-oriented data security approach is considered more robust, secure, and efficient because quantum deals with an infinite number of potential states along with zero and one state whereas classical considers either zero or one as the possible outcome states. Various quantum gate permits a rotational outcome in a 360$^{\circ}$ view that analyzes the input data deeply with numerous possible qubit states to generate possible outcomes from it and thus predicts the data breach more adequately. 
\subsection{Key Contributions}\label{subseccontri}
In light of the aforementioned approaches, the fivefold key contributions of this paper are discussed below. 
\begin{enumerate}
	\item{A novel quantum driven IQ-HDM framework using the computational efficiency of \textit{quantum encryption} and \textit{quantum feed-forward neural network} approaches is designed to furnish end-to-end management of healthcare data ensuring secure data storage, efficient data communication, and prediction of malicious entities.}
		\item{The \textit{QOTPE} unit is designed that is responsible for encryption of data in the form of quantum states resulting in maximally mixed states, providing a perfect and unconditional security to the transmitted data.}
	\item{A \textit{QPHDC} unit incorporating pauliX, Hadamard quantum gates, and qubits is developed that eventually add more potency to data communication by allowing the secure sharing of data among various stakeholders.}
	\item {The proposed framework strengthens data communication by proactively mitigating the hazardous data request intentions and recognizing the malicious entity to prevent further breaches.}
	\item{A series of experiments are conducted utilizing the widely adopted four benchmark datasets that demonstrate the efficacy of the proposed end-to-end quantum-oriented approach for improving the security of electronic healthcare data management. The accomplished results are compared with the state-of-the-art works through diverse performance metrics.}
\end{enumerate}
\subsection{Paper Outline}\label{subsecoutline}
This article is structured as follows. Section \ref{secint} discusses introduction and related work with key contributions of the presented research work. Section \ref{secIQ} furnishes a detailed elaboration of the proposed IQ-HDM framework involving two units QOTPE and OPHDC, to ensure comprehensive data management as explained in Section \ref{secQOTPE} and Section \ref{secprediction}, respectively. The design and complexity of IQ-HDM are conferred in Section \ref{secopd}. The performance evaluation followed by discussion remarks about the proposed work is presented in Section \ref{secperformance}. The conclusive remarks and the future scope of the proposed work are outlined in Section \ref{seccon}. 
 Table \ref{tab:tabnotation} shows the list of symbols with explanatory terms used throughout this article.
 \begin{table}[!htbp]
	\caption{Nomenclature}
	\label{tab:tabnotation}
	\begin{center}
		\resizebox{0.95\columnwidth}{!}{
			\tiny
			\begin{tabular}{llll}\hline 
				$\mathbb{H}^a$ &Healthcare agency & $\mathbb{DU}$ &Data users \\
				$\mathbb{T}^p$ &Third party & $\mathbb{CSS}$ &Cloud Service Supplier \\
				$m$ & Number of users & $D^h$ & Healthcare data objects \\
				$\mathbb{DSS}$ & Data Storage Server & $\vert \psi \rangle$ & Quantum states \\
				$\mathbb{R}_d$ & Quantum superposition  &	$\mathcal{DB}^{grand}$ & Total number of\\
				 &  of bit strings& &data access \\
				$\mathbb{R}_i$ & Data access request &	$\mathbb{DU}_i^{\star}$ & Live details \\
				$\mathbb{DU}_i^{\star\star}$ &  Historical details  &	$\rho$ & Breach susceptibility  \\
				$\xi$ & Eligibility parameters &	$\mathcal{AU}$ & Users’ authenticity \\
				$\mathcal{AD}$ & Authorized data &	$\mathcal{RF}$ & Risk factor \\
				$t_{a^\ast}$, $t_{b^\ast}$ & Time-interval &	$\phi$ & Past leakage status \\
				$\mathcal{DB}^{mal}$ & Malicious data breach &	$\Pi$ & Breach factor \\
			 $\alpha, \beta$ & Padding keys &	$\doublecap$ & Accuracy \\
				\hline
		\end{tabular}}
	\end{center}
\end{table}
\section{IQ-HDM}\label{secIQ}
\begin{figure*}[!ht]
\centering
\includegraphics[width=\linewidth]{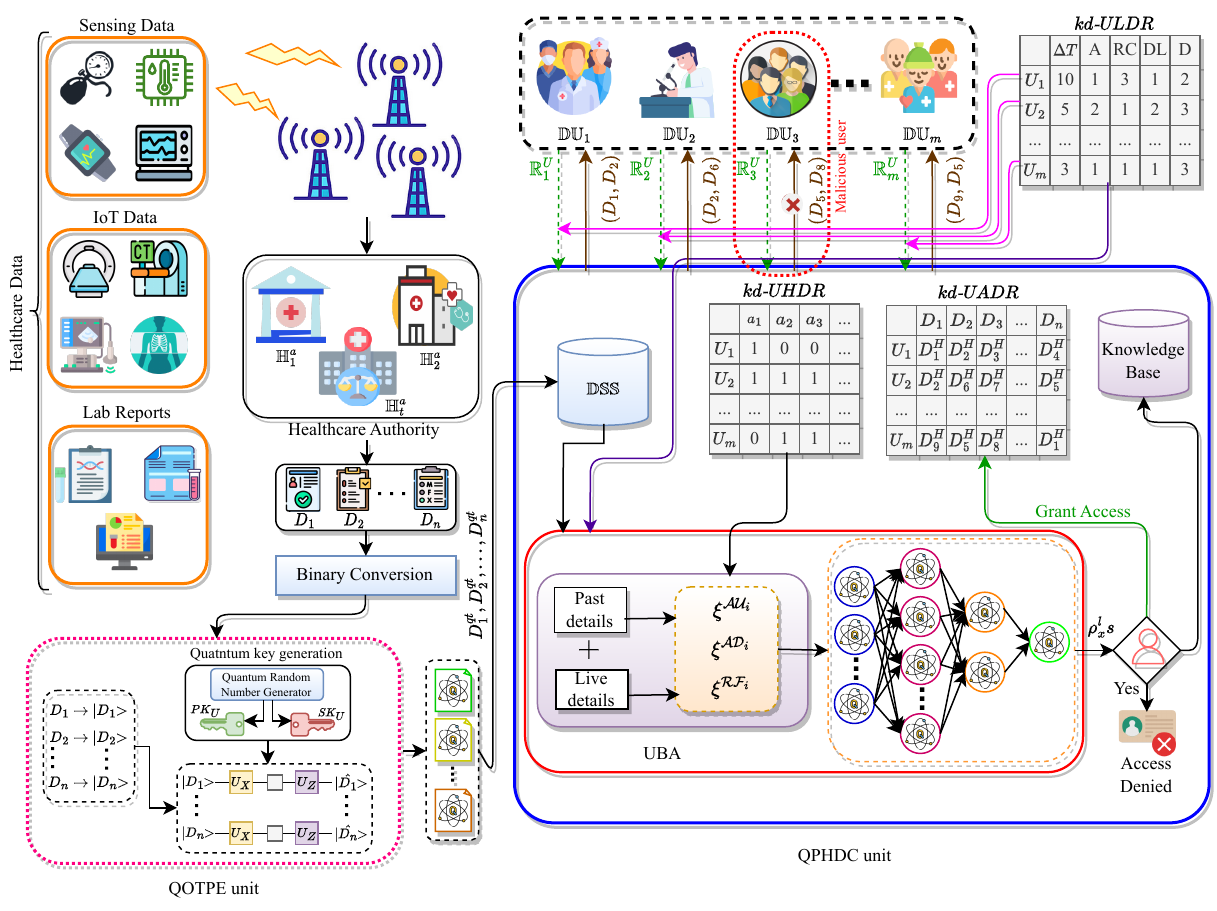}
\caption{Architecture of the proposed intelligent quantum oriented healthcare data management framework.}
\label{fig1}
\end{figure*}

This section describes the framework entities and their designated roles, and summarizes the workflow of IQ-HDM. The comprehensive architecture of the proposed framework is depicted in Fig. \ref{fig1}.
\subsection{System Model}\label{subsecsysmod}
The system model comprises of four entities \textit{healthcare agencies} ($\mathbb{H}^a$), \textit{cloud service supplier} ($\mathbb{CSS}$), \textit{data users} ($\mathbb{DU}$), and \textit{third parties} ($\mathbb{T}^p$) which are defined as follows.
\begin{enumerate}
\item{\textit{Healthcare Agencies} ($\mathbb{H}^a$): An entity generating healthcare data using various Sensors and IoT devices. $\mathbb{H}^a$ treats $\mathbb{CSS}$ as trusted but is curious-to-know therefore it encrypts the data before transferring it for storage and sharing purposes. Moreover, $\mathbb{H}^a_i$ itself might not leak its data, but may leak the other $\mathbb{H}^a_j$’s data, therefore is considered an untrusted entity.}
\item{\textit{Cloud Service Supplier} ($\mathbb{CSS}$): An entity that collects all the encrypted data from $\mathbb{H}^a_i$ to offer storage, computation for further sharing among $\mathbb{DU}$. $\mathbb{CSS}$ supports secure data communication and malicious entity estimation by deploying a QFNN based quantum-protected healthcare data communication unit (QPHDC).}
\item{\textit{Data User} ($\mathbb{DU}$): An entity raising a request to $\mathbb{CSS}$ for grant of \textit{healthcare data objects} ($D^h$) access required for different utility purposes and obtains quantum encrypted $D^h$ along with the key. $\mathbb{DU}_i$ is considered a highly untrusted entity.}
\item{\textit{Third Party} ($\mathbb{T}^p$): An unauthorized and untrusted entity that belongs indirectly to the system. $\mathbb{T}^p$ can access the relevant information from an malicious entity or by accessing $D^h$ illegally from some authorized entity.}
\end{enumerate}

Considerable \textit{healthcare agencies} $\mathbb{H}^a$ are accumulating crucial medical data from different sources such as sensor devices, IoT devices, and diagnosis reports, etc. as displayed in Fig. \ref{fig1}. $\mathbb{H}^a$, share \textit{healthcare data objects} $D^H_i$ with multiple \textit{data users} $\mathbb{DU}_i$ utilizing data to deliver better medical services. Data sharing encounters the threat of safety, privacy, and breaches. The proposed framework considers entities $\mathbb{H}^a$, $\mathbb{DU}$, and $\mathbb{T}^p$ are untrusted and $\mathbb{CSS}$ is semi-trusted but curious-to-know entity. Specific challenges are described as follows.
\begin{itemize}
\item{Lack of secure data storage at cloud premises due to the possibility of being misused by curious-to-know $\mathbb{CSS}$.}
\item{Assessment of intention for data request before $D^h_i$ is granted because $\mathbb{DU}_i$ might be mal-intentional.}
\item{Proactive estimation of malicious user might responsible for unauthorized transmission of sensitive $D^h_i$ to particular $\mathbb{T}^p$.}
\end{itemize}
To ensure data security during transmission and to keep data hidden from curious-to-know cloud service supplier ($\mathbb{CSS}$), data owner $\mathbb{H}^a$ encrypts their respective data into some directly unreadable format by utilizing the QOTPE approach. Also, QFNN oriented QPHDC unit is employed for proactive estimation of potential breach and malicious entity by performing extensive analysis of each user for each data access request. The purpose of IQ-HDM is to anticipate an advanced, suitable quantum-driven solution for the overall management of crucial healthcare data which delivers secure storage, mitigates communication issues, and predicts malicious entity in case of data leakage by utilizing extremely powerful quantum approaches imparting high privacy, robust security and, mitigating the threats, to intensify the overall performance of the system.
\subsection{Quantum Encryption for Outsourced Data}\label{secQOTPE}
To make data secure before allowing its transmission to a shared cloud platform data is encrypted into some unreadable content by deploying the quantum one-time padding encryption (QOTPE) approach. 
The encrypted data is then stored at a cloud data storage server ($\mathbb{DSS}$) thus enabling highly secure data storage. It comprises following consecutive steps: 
%
\subsubsection{Encoding}
In order to perform any quantum computations, the classical data needs to be converted to the quantum states, which is achieved through basis encoding in the proposed model. The classical data is converted to the equivalent binary strings as$({D^H_1})_2=, (D^H_2)_2=b_0 b_1 b_{x-1} b_x, \cdots, (D^H_n)_2=n_0 n_1 n_{x-1} n_x$, where $a, b,\cdots, n$ subscripts denote the individual binary digits for each data instance. Thereafter, each classical bit of data instance $D_i^H$ is encoded as quantum state $\vert \psi^a_0\rangle \vert \psi^a_1\rangle \cdots \vert \psi^a_{x-1}\rangle \vert \psi^a_x\rangle, \vert \psi^b_0\rangle\vert \psi^b_1\rangle \cdots \vert \psi^b_{x-1}\rangle \vert \psi^b_x\rangle,  \cdots,  \vert \psi^n_0\rangle$ $\vert \psi^n_1 \rangle \cdots \vert \psi^n_{x-1}\rangle \vert \psi^n_x\rangle$ by initializing corresponding quantum registers $QR_i$, $\forall i=1, 2, \cdots, n$, defined in Eq. (\ref{Eq:QR}), along with application of Controlled-Not gate on required qubits, given in Eq. (\ref{Eq:U}).
\begin{align} \label{Eq:QR}
	QR = \{\vert \psi_1\rangle, \vert \psi_2 \rangle, \ldots, \vert \psi_n\rangle\}
\end{align}
\begin{align} \label{Eq:U}
	\begin{split}
	U(\vert \psi_i \rangle) = \text{CNOT}(\vert 1 \rangle, \vert \psi_i \rangle) \cdot \vert 1 \rangle \langle 1 \vert + \\ \text{Id}(\vert 0 \rangle, \vert \psi_i \rangle) \cdot \vert 0 \rangle \langle 0 \vert \cdot \vert \psi_i \rangle
	\end{split}
\end{align}
where, $U(\vert \psi_i \rangle)$ denotes the application of unitary gate to the $ith$ input data state $\vert \psi_i \rangle$. $\text{CNOT}(\vert 1 \rangle, \vert \psi_i \rangle)$ representing the application of CNOT gate to the target qubit $\vert \psi_i \rangle$ along with the control qubit $\vert 1 \rangle$. $\text{Id}(\vert 0 \rangle, \vert \psi \rangle)$ is the identity gate applied to the target qubit $\vert \psi_i \rangle$ and the control qubit $\vert 0 \rangle$. Furthermore, the projectors being applied over the states $\vert 1 \rangle$ and $\vert 0 \rangle$ are $\vert 1 \rangle \langle 1 \vert$ and $\vert 0 \rangle \langle 0 \vert$, respectively. The resultant quantum states ($\vert \psi_i\rangle$) are capable of performing multiple computations simultaneously due to underlying quantum mechanical properties such as entanglement and superposition. A state vector with $m$ number of qubits used for the precision and $d$ number of samples, is represented as $m+[log(d)]$ and $x = (x_1, . . . , x_d)$ $\in$ $\mathbb{R}_d$ denoted as quantum superposition of bit strings, in which each instance is a binary string formed using $N$ bits for the basis encoding. Furthermore, for $x_i = (b_1, . . . , b_j, . . . , b_N)$ for $j = 1, . . . , N$ with $b_j$ $\in$ $\{0,1\}$, basis encoding is stated in Eq. (\ref{eq1}).
\begin{equation}
	\label{eq1}
	\vert x \rangle= \frac{1}{\sqrt{d}} \sum_{i=1}^{d}  \vert x_i \rangle
\end{equation}
Following the encoding of classical data, in order to achieve a secure communication, the equivalent quantum states need to be encrypted before being transmitted to the cloud, which is accomplished by utilizing quantum one time padding as described in forthcoming subsection.
\subsubsection{Encryption}
\begin{figure}[!htbp]
    \centering
    \includegraphics[width=0.90\linewidth]{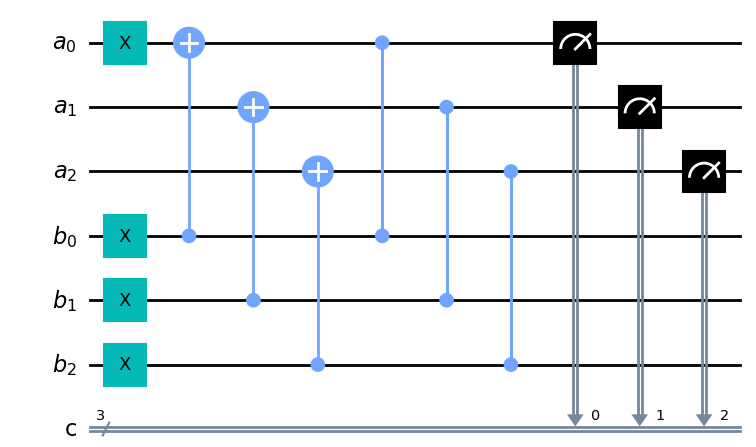}
    \caption{Circuit for Quantum one time padding with 3 qubits}
    \label{fig:QOTP}
\end{figure}

The quantum-mechanical principles underlying in the quantum states establish the information-theoretical security of the quantum information. The quantum analogy of classical one time pad is quantum one time pad that is one of the most examined quantum encryption algorithms.   
The security of quantum information is established on the principles of quantum mechanics, which is information-theoretically-secure. QOTPE is amongst the most investigated techniques in quantum encryption \cite{cheng2019batten}. For each quantum state $\vert \psi\rangle$, two randomly generated keys $\alpha, \beta \in \{0,1\}^n$ are used for padding of original information. For secret-key quantum encryption, it is assumed that secret keys are known to both receiver and sender. This bit-wise quantum one time pad protocol can be represented as $X^{\alpha}$= $\otimes_{i=1}^{n} \sigma_x^{\alpha(i)}$ and $Z^{\beta}$= $\otimes_{i=1}^{n} \sigma_z^{\beta(i)}$, where $\sigma$ is the operation performed over the given qubit. Corresponding to $X^{\alpha}$, $\sigma_x$ is applied to the bits at positions in the n-bit string $\alpha$ and analogously for $Z^{\beta}$, which leads to a maximally mixed state that is completely unidentifiable for the attacker. The resultant encrypted state in generalized form can be rewritten as stated in Eq. (\ref{eq2}).
\begin{align} \label{eq2}
    \begin{split}
    Enc_{\psi}= & \otimes_{i=1}^{n}X^{\alpha_i}Z^{\beta_i}\vert\phi_i\rangle\\
    =& X^{\alpha_1}Z^{\beta_1}\otimes X^{\alpha_2}Z^{\beta_2}\otimes...\otimes X^{\alpha_n}Z^{\beta_n} \\
    & \vert\phi_1\rangle\vert\phi_2\rangle ... \vert\phi_n\rangle   
    \end{split}
\end{align}
The circuit generated for quantum one time padding is provided in Fig. \ref{fig:QOTP} which is tested using six qubits. It depicts the QOTPE circuit equivalent to Eq. (\ref{eq2}) but the encryption keys are not presented in the circuit as the circuit is generated on the quantum computer; where encryption keys are being used in back-end. The maximally mixed states are retrieved after encryption and subsequently delivered to the cloud securely. Eq. (\ref{eq3}) demonstrates the perfect security achieved by QOTPE, which is as follows:
\begin{align} \label{eq3}
\Tilde{\rho}=& \sum_{k=1}^{M}p_kU_k\rho U_k^{\dagger} \nonumber\\
    =& \frac{1}{2^{2n}}\sum_{\alpha, \beta \in \{0,1\}^n}X^{\alpha}Z^{\beta}\rho(X^{\alpha}Z^{\beta})^{\dagger} = \frac{I_{2^n}}{2^n}
\end{align}
where $\rho$ denoting the data quantum state and the maximally mixed state for $n$ qubits is represented by $\frac{I_{2^n}}{2^n}$.

\textit{Security Proof}: The most general scheme to form an encryption framework for any n-qubit system is to have a set of M operations $\{U_k\}, k=1,2,..., M$, where each element $U_k$ is a $2^n \times 2^n$ unitary matrix. Any random number being used as key $k$ with probability $p_k$ and each quantum state is encrypted through the corresponding unitary operation $U_k$. Subsequently, decryption is achieved through application of $U_k^{\dagger}$  to obtain the actual state. Considering $\rho$ as the input state and $\Tilde{\rho}$ is the equivalent encrypted state. In order to have the protocol to be informationally secure, every output state $\Tilde{\rho}$ must be a maximally mixed state, corresponding to each input state $\rho$. Therefore, to prove the perfect security of the quantum one time pad protocol, we consider $p_k=1/2^{2n}$ and $U_k=X^{\alpha}Z^{\beta}$, where $\alpha, \beta \in \{0,1\}^n$. The inner product of two matrices $M_1$ and $M_2$ is defined as $Tr(M_1, M_2^{\dagger})$. Furthermore, considering a set of $2^n \times 2^n$ matrices as an inner product space, it is trivially verifiable that the set of $2^{2n}$ unitary matrices $\{X^{\alpha}Z^{\beta}\}$ results in an orthonormal basis. If any input message $\rho$ is expanded in the $X^{\alpha}Z^{\beta}$ basis, is retrieved as in Eq. (\ref{eq4}).
\begin{equation}\label{eq4}
    \rho= \sum_{\alpha, \beta} A_{\alpha, \beta}X^{\alpha}Z^{\beta}, 
\end{equation}
where, $A_{\alpha, \beta}$ is equivalent to $Tr(\rho Z^{\beta}X^{\alpha})/2^n$. Thereby, the perfect security for underlying protocol is given by satisfying maximally mixed state through Eq. (\ref{eq5}).

\begin{gather} \label{eq5}
\begin{split}
    \sum_{k=1}^{M}p_kU_k\rho U_k^{\dagger}=&\frac{1}{2^{2n}} \sum_{\zeta, \eta} X^{\zeta}Z^{\eta}\rho Z^{\eta}X^{\zeta} \\
    =& \frac{1}{2^{2n}} \sum_{\alpha, \beta} A_{\alpha, \beta}\sum_{\zeta, \eta}X^{\zeta}Z^{\eta}X^{\alpha}Z^{\beta}Z^{\eta}X^{\zeta} \\
    =& \frac{1}{2^{2n}} \sum_{\alpha, \beta} A_{\alpha, \beta}\sum_{\zeta, \eta} (-1)^{\alpha \eta \oplus \zeta \beta} X^{\alpha}Z^{\beta}\\
    =& \sum_{\alpha, \beta} A_{\alpha, \beta} \eta_{\alpha,0} \eta_{\beta,0}X^{\alpha}Z^{\beta}\\
    =& A_{0,0}I= \frac{Tr(\rho)}{2^n}= \frac{1}{2^n}I
    \end{split}
\end{gather}
\subsection{Quantum Prediction for Malicious Entities}\label{secprediction}
Let us assume $m$ \textit{data users}: $\{\mathbb{DU}_1, \mathbb{DU}_2, ...,  \mathbb{DU}_m\} \in \mathbb{DU}$ raises the request $\{\mathbb{R}_1, \mathbb{R}_2, ..., \mathbb{R}_m\}$ to achieve access for sensitive data $D^h_i$. Each $[\mathbb{R}_i: \langle D^h_i, \mathbb{DU}_i^{\star}\rangle]$ comprising details for required $D^h_i$ and requesting user's current details such as type of data, amount of data, request channel etc. are provided to $\mathbb{CSS}$ for further analysis to determine the intention of user behind the data access request $\mathbb{R}_i$. $\mathbb{CSS}$ employs quantum feed forward neural-network (QFNN) based module to proactively determine possible malicious user by extensive analysis of live details $\mathbb{DU}_i^{\star}$ supplied with request from  (\textit{knowledge database-users' live details repository} (kd-ULDR) and historical details $\mathbb{DU}_i^{\star\star}$ such as leakage record, leakage channel etc. accessible from \textit{knowledge database-users' historical details repository} (kd-UHDR). $\mathbb{DU}_i^h$ are considered to be \{\textit{known}, \textit{unknown}, \textit{mal-intentional}\}. 
\begin{figure*}[!htbp]
	\centering
	\includegraphics[width=0.99\linewidth]{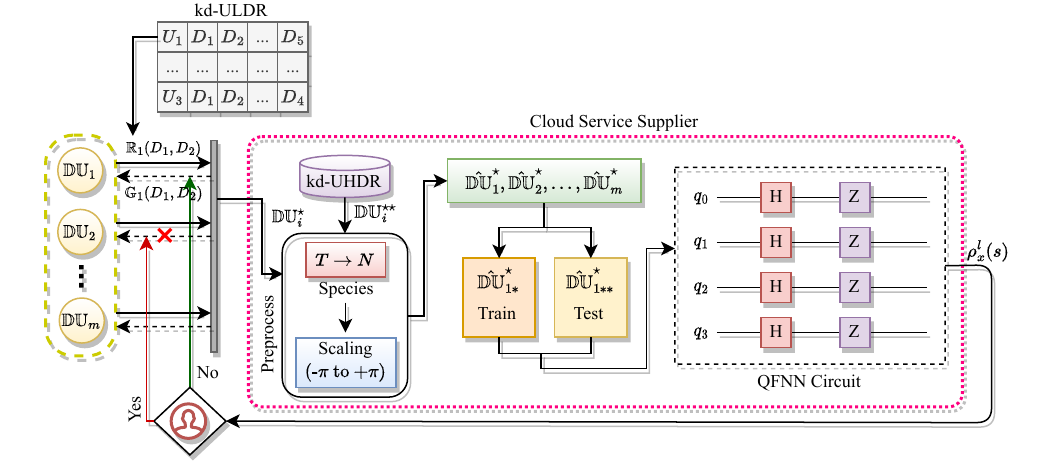}
	\caption{QFNN based QPHDC to predict malicious entity for secure data communication.}
	\label{figpred}
\end{figure*}
\subsubsection{User Behaviour Analysis}
The user intention for being malicious or non-malicious, is evaluated prior to data access grant. Eqs. (\ref{final1} and \ref{final2}) assesses the breach susceptibility ($\rho$) of the {$\mathbb{DU}_i$}, established on basis of ${N^\ast}$ user eligibility parameters $\xi$ such as \textit{users' authenticity} ($\mathcal{AU}$); \textit{authorized data} ($\mathcal{AD}$); \textit{risk factor} as computed in Eqs. (\ref{rp1}-\ref{rp8}).
\begin{gather}
\label{final1}
\xi =\xi^{\mathcal{AU}_i} +\xi^{\mathcal{AD}_i}+\xi^{\mathcal{RF}_{i}} + \xi^{N^\ast}\\
\label{final2}
    \rho^{\mathbb{DU}_i}  = 
      \begin{cases}
           \textit{Non-malicious}\; (1),  & \text{If} (\mathcal{\xi} < 1)\\
            \textit{Malicious} \;(0), & Otherwise\\
        \end{cases}
        \end{gather}
\noindent {\textit{Users' authenticity ($\mathcal{AU}$):}} User is validated using Eq. (\ref{rp1}) through login credential ($\mathbb{LC}$). Eq. (\ref{rp2}) determine whether the user $\mathbb{DU}_i$ is `existing' or `new'.
\begin{gather} \label{rp1}
\mathcal{AU}_i = 
 \begin{cases}
 \textit{Authentic},  & \text{If} (\mathbb{LC} \cup \mathbb{RA} = match) \\
 \textit{Un-authentic}, & Otherwise\\
 \end{cases} \\
 \label{rp2}
\mathcal{\xi}^\mathcal{AU}_i = 
 \begin{cases}
  \textit{Existing} \; (0),  & \text{If}  (|\Im_i| > 0)\\
 \textit{New} \; (1), & Otherwise\\
 \end{cases}
 \end{gather}
\noindent {\textit{Authorised data ($\mathcal{AD}$):}} Every user is allowed to raise request $\Re_i$ as computed in Eq. (\ref{rp3}) and Eq. (\ref{rp4}), for predefined set of data ($\mathcal{AD}_i$) only, for example a patient can only access personal data, not the entire healthcare data. Here, $z_{1}$, $z_{2}$, $\dots$, $z_{m^\ast}$ specifies the number of datasets from different medical categories: $w_1$, $w_2$, $\dots$, $w_{m^\ast}$, respectively. 
\begin{align}
	\begin{split}
		\label{rp3}
		\mathcal{AD}_i = (w_1 \times \sum_{k=1}^{z_{1}}D_{k}) \cup (w_2 \times \sum_{k=1}^{z_{2}}D_{k}) \, \cup \\
		 \dots \cup (w_{m^\ast} \times \sum_{k=1}^{z_{m^\ast}}D_{k})
	\end{split}	
\end{align}
\begin{equation}\
  \label{rp4}
\mathcal{\xi}^{\mathcal{AD}_i} = 
 \begin{cases}
 \textit{Legal} \; (1),  & \text{If} (\Re_{i} \times (w_i \times D_{i}) \subseteq \mathcal{AD}_{i})\\
  \textit{Illegal} \; (0), & Otherwise\\
 \end{cases}
 \end{equation}
\noindent {\textit{Risk factor ($\mathcal{RF}$):}} Suppose the associated user $\mathbb{DU}_i$ has demanded data \{$D^h_1$, $D^h_2$, $\dots$, $D^h_n$\} during time-interval \{$t_{a^\ast}$, $t_{b^\ast}$\} and the status ($\phi$) for any past leakage is stated in Eq. (\ref{rp5}). The total number of $\mathcal{DB}^{mal}$ during this period is estimated using Eq. (\ref{rp6}). The breach factor ($\Pi$) by $u_{i}$ is computed using Eq. (\ref{rp7}), where $\mathcal{DB}^{grand}$ is total number of data access over period \{$t_{a^\ast}$, $t_{b^\ast}$\}. The data breaches frequency is computed in Eq. (\ref{rp8}) where $\sum\limits_{k=1}^{H} Dz_{k} \not\in \mathcal{AD}_{i}$ and $t_{ijk}$ represents number of times $u_{i}$ has endeavored to access unauthorized data ($Dz_{k}$) over $j^{th}$ time-period. The term $H$ and $M$ stands for total number of unauthorized data requested by $u_{i}$ during time duration $M$ where, $M$ $\in$ \{$t_{a^\ast}$, $t_{b^\ast}$\}.
\begin{gather} \label{rp5}
{\phi_i} = 
 \begin{cases}
  \textit{True (1)},  & \text{If} (Breach = yes)\\
 \textit{False (0)}, & Otherwise\\
 \end{cases} \\
\label{rp6}
    \mathcal{DB}_i^{mal} = \sum_{i=1}^{z}(D_{i} \times \phi_{i} \times t) \quad  \forall_t \in \{t_{a^\ast}, t_{b^\ast}\} \\
\label{rp7}
\int_{t_{a^\ast}}^{t_{b^\ast}} \Pi_{i} dt = \int_{t_{a^\ast}}^{t_{b^\ast}} \frac{\mathcal{DB}_i^{mal}}{\mathcal{DB}_i^{grand}} dt \\
\label{rp8}
    \mathcal{FDB}^{mal}_{i} = |\sum_{k=1}^{H}\sum_{j=1}^{M} Dz_{k} \times t_{ijk} \times u_{i}|
\end{gather}   
Eq. (\ref{rp9}) computes the risk factor ($\mathcal{RF}$) associated with a $\Re_i$ and Eq. (\ref{rp10}) determines whether the data demand should allow utilizing cloud services for data access or not.
\begin{gather}
\label{rp9}
    \mathcal{RF}_i = \Pi_i \times \mathcal{FDB}^{mal}_{i} \\
\label{rp10}
\xi^{\mathcal{RF}_{i}} = 
 \begin{cases}
  \textit{insensitve} \; (0),  & \text{If} (Thr^{risk} > \mathcal{RF}_{i})\\
  \textit{sensitve} \; (1), & Otherwise\\
 \end{cases}
 \end{gather}
\subsubsection{Malicious Entity Prediction}
Before the requested data is granted to $\mathbb{DU}_i$, a proactive analysis of the intention behind the request is performed to safeguard the communication. Only after being recognized as a legitimate user, requested data is granted to the authorized user and \textit{users' allocated data repository} (kd-UADR) is updated, accordingly. As shown in Fig. \ref{figpred} a \textit{quantum feed-forward neural network} machine learning-based algorithm is employed to accomplish proactive estimation of the malicious user. The key idea of the QFNN-based QPHDC unit is to optimize the user's request parameters according to the cost function as illustrated in Eq. (\ref{eq7}). Moreover, \text{cost function} assumes to be given \text{training data} and \text{output states} to compute the cost function as stated in Eqs. (\ref{eqtd} and \ref{eq7}), respectively. 
\begin{align}\label{eqtd}
    \begin{split}
    trainingData[x][1] = \vert \phi_x^{out} \rangle \\
    outputStates[x] = \vert \rho_x^{out}(s) \rangle 
\end{split}
\end{align}
\begin{equation}\label{eq7}
 C(s) = \frac{1}{N} \sum_{x = 1}^N \vert \phi_x^{out} \rangle \rho_x^{out}(s) \phi_x^{out}
\end{equation}

\begin{figure}[!ht]
\centering
\includegraphics[width=\linewidth]{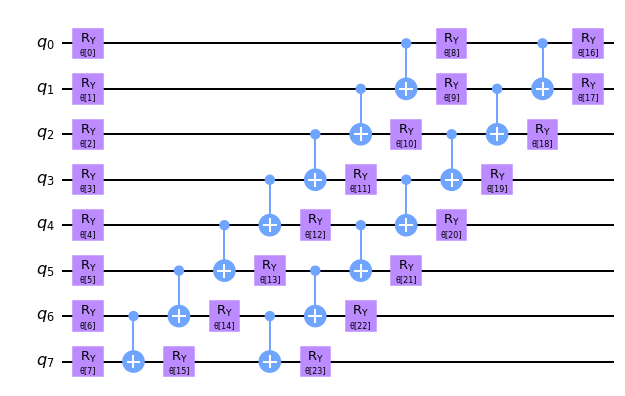}
\caption{Quantum circuit for qubit embedding}
\label{figquant}
\end{figure}

QFNN architecture creates \text{adjoint layer channel} that can be described as a 4-tuple trainable quantum neural network like (\text{QFNN architecture}, \text{unitaries}, \text{training Data}, \text{network unitary}) as represented in Eq. (\ref{eq20}). Fig. \ref{figquant} is displaying the quantum circuit for qubit embedding in QFNN prediction unit.
\begin{align}\label{eq20}
    \begin{split}
        \mathcal{F}_s^l(X^{l}) &= tr_{l}\bigg( \big( \mathbb{1}_{l-1} \otimes 0 \dots 0_l 0 \dots 0 \big) \\ 
        & U^{l}(s)^{\dagger} \big( \mathbb{1}_{l-1} \otimes X^l \big) U^l(s) \bigg) \\
        =& tr_{l}\bigg( \big( \mathbb{1}_{l-1} \otimes 0 \dots 0_l 0 \dots 0 \big) \ U_1^l(s)^{\dagger} \dots U_{m_l}^l(s)^{\dagger} \\
        & \big(\mathbb{1}_{l-1} \otimes X^l \big) U_{m_l}^l(s) \dots U_1^l(s)\bigg)
    \end{split}
\end{align}
for \text{input state} = $X^{l}$. 

Feed-forward neural network assumed to be given QFNN architecture, unitaries, and training data as usual, to carry out the malicious user estimation task as described in efficient learning for deep quantum neural networks in the following steps:
\begin{itemize}
    \item For each element $\big[ \phi_x^{in}, \phi_x^{out} \big]$ in \text{training data} do:
    \item Calculate the network input 
        $\rho_x^{in} = \phi_x^{in} \phi_x^{in}$
    \item For every layer $l$ in \text{QFNN architecture} do:
    \item Apply the layer channel $\mathcal{E}_s^l$ to the output of the previous layer $l-1$
    \item Store the result $\rho_x^l(s)$
\end{itemize}
The probability of \text{cost Function} returns the average as computed in Eq. (\ref{eq21}).
\begin{equation}\label{eq21}
    \mathrm{Pr}\l(\tilde{m}(\Phi)\neq y| m(\Phi)=y) \approx \sigma \left(\sqrt{R} \frac{\frac{1-yb}{2}-\hat{p}_y}{\sqrt{2(1-\hat{p}_y)\hat{p}_y}}\right)
\end{equation}
\section{Operational Design and Complexity}\label{secopd}
Algorithm \ref{algo-osecc} imparts the operational summary of IQ-HDM, by utilizing the computational efficiency of quantum computing to provide shielded data storage and determines data breaches proactively for secure healthcare communications in the distributed cloud environment.
	\begin{algorithm}[!htbp]
		\caption{IQ-HDM: Operational Summary}
		\label{algo-osecc}
        \textbf{Input}: Knowledge databases including \textit{User Historical Data Repository (kd-UHDR)} and \textit{User Allocated Data Repository (kd-UADR)}\; 
        
        \textbf{Initialize}: Data request, user details and related attributes \;
        Perform Quantum Encryption, to shield the data in storage \;
        \For{each data object $\{D^{\mathbb{H}^a_1}, D^{\mathbb{H}^a_2}, ..., D^{\mathbb{H}^a_p}\}$} {QE using \textit{QOTPE} as computed in Eq. (\ref{eq2}) \;}
		Periodical, training of \textit{QPHDC} for secure communication \;
		\For {every data user $\{\mathbb{DU}_1, \mathbb{DU}_2, ...,  \mathbb{DU}_m$\}}{
		\For{each data request ($\{\mathbb{R}_1, \mathbb{R}_2, ..., \mathbb{R}_m\}$ \{1, 2, ..., m\} from respective data user ($\mathbb{DU}$)}{
		Examine the probable purpose of data request by \textit{QFNN}- qubit measurement as computed in Eq. (\ref{eq7})\; 
		\eIf{malicious} {Request $\mathbb{R}_i$ leads to 'Data Breach'\;}{Grant data access\;}
	}
	}
	\end{algorithm}

\textit{Time complexity}: Steps (1-2) performs basic operations such as input the required datasets and initializes a data request, user details, and related attributes, consuming $\mathcal{O}(1)$ complexity, while steps (3-6) involves the encoding, encryption of crucial data before storage on shared data cloud platform, contributing complexity $\mathcal{O}(QE)$. The periodical training of QFNN predictor where the complexity depends on the quantum gates and circuits rendering $\mathcal{O}(\Tilde{Q})$. Steps (8-14) iterate for $m$ users, wherein steps (9-15) replicate for $m$ data request. Step 10 examines the probable purpose of data request by deploying QFNN-based QPHDC, to find the users' intentions for being malicious or non-malicious show complexity $\mathcal{O}({N^{\ast}})$. Steps (11-15) grant or deny healthcare data access depending upon the anticipated intention of the data request inducing $\mathcal{O}(1)$ complexity. Hence, the absolute complexity comes out to be $\mathcal{O}(n\times QE \times \Tilde{Q} \times N^{\ast}) \Rightarrow$ $\mathcal{O}(nQE\Tilde{Q}N^{\ast})$.
\section{Performance Evaluation}\label{secperformance}
\subsection{Experimental Setup and Implementation}\label{subsecexp}
The experimental work is carried out on a server machine encompassing two Intel\textsuperscript{\textregistered} Xeon\textsuperscript{\textregistered} Silver 4114 CPU with a 40 core processor and having 2.20 GHz clock speed. The simulation machine run on Ubuntu 16.04, an 64-bit LTS operating system comprising 128 GB of main memory RAM. Enactment of proposed work is carried out using Python 3.9. Moreover, IQ-HDM is simulated using the IBM Qiskit platform (version 0.43.0). The simulation is conducted on IBM QASM simulator and IBM quantum systems including IBM Nairobi (No. of qubits used-7), IBM Perth (No. of qubits used-5), selected on the basis of availability of the system along with most suitable parameters such as number of qubits supported by the system, quantum volume and number of jobs being queued. Also, QFNN to carry out prediction work employs Adam optimizer on two qubits and four qubits by utilizing Hadamard and CNOT quantum gates. However, due to execution constraints of available quantum computer instances and classical simulator, small chunks of dataset are used to run the experiments. Performance of framework under consideration, is examined through a dataset comprising of 10k agents live details alongwith ancient details. Major live details parameters are type of profession, number of requests from agent, type of requests from agent, and data limit for which data was accessed whereas the major ancient details parameters are ancient data of agents, leaked or never leaked data, how many times leaked the data, how frequently asking for data, and data retention. These agents all together are classified into three strictly different brackets which are non-malevolent, malevolent, and unknown. Moreover, framework assumes all the entities as non trusted to carry out execution task. The important primitives related to execution over quantum computer are listed in Table \ref{tab:QC}.
\begin{table}[!htbp]
	\caption{Experimental Setup Parameters for the Quantum Computers used and their values}
	\setlength{\tabcolsep}{3pt}
	\begin{tabular}{p{130pt}|p{50pt}|p{50pt}}
		\hline
		\textbf{Quantum Computer Used}& \textbf{IBM Perth} & \textbf{IBM Nairobi} \\
		\hline \hline
		Version &1.2.8&1.3.3 \\	
		Number of Qubits & 7&5,7\\
		Quantum Register Size &7 &7\\
		Classical Register Size &2 &3 \\
		Number of Shots &300 & 300 \\
		Execution Time on Quantum Computer &1.987 (s)&6.41 (s) \\
		Total Execution Time & 1h:14m:24s & 0h:49m:42s \\
		\hline
	\end{tabular}
	\label{tab:QC}
\end{table}
\subsection{Datasets and Simulation Parameters}\label{subsecdata}
IQ-HDM is evaluated using different benchmark datasets available in public real workload datasets. For quantum encryption purposes, the following datasets are employed: 1) Covid-19 surveillance \cite{covid-19_surveillance_567}, 2) TCGA \cite{glioma_grading_clinical_and_mutation_features_759}, and 3) Diabetes \cite{early_stage_diabetes_risk_prediction_dataset._529}. COVID-19 surveillance data categorizes the health details into three categories based on seven different health parameters. TCGA data is comprised of 839 instances with twenty-three features related to Gliomas, the most common primary tumors of the brain. This dataset considers, the most frequently mutated 20 genes and 3 clinical features from TCGA-LGG and TCGA-GBM brain glioma projects to determine whether a patient is LGG or GBM. Diabetes provides information gathered by monitoring sixteen health parameters such as Age, Gender, Polyuria, Polydipsia, Sudden weight loss, Weakness, Polyphagia, Genital thrush, Visual blurring, Itching, Irritability, Delayed healing, Partial paresis, Muscle stiffness, Alopecia, Obesity, for a set of 520 patients, to identify whether the patient diabetes is positive or negative. In this context, for malicious entity prediction by QFNN, an extended version of CMU CERT synthetic insider threat dataset r4.2 \cite{dataset} is employed. 
\subsection{Computational Analysis}
\subsubsection{QOTPE Result}\label{subsecenc}
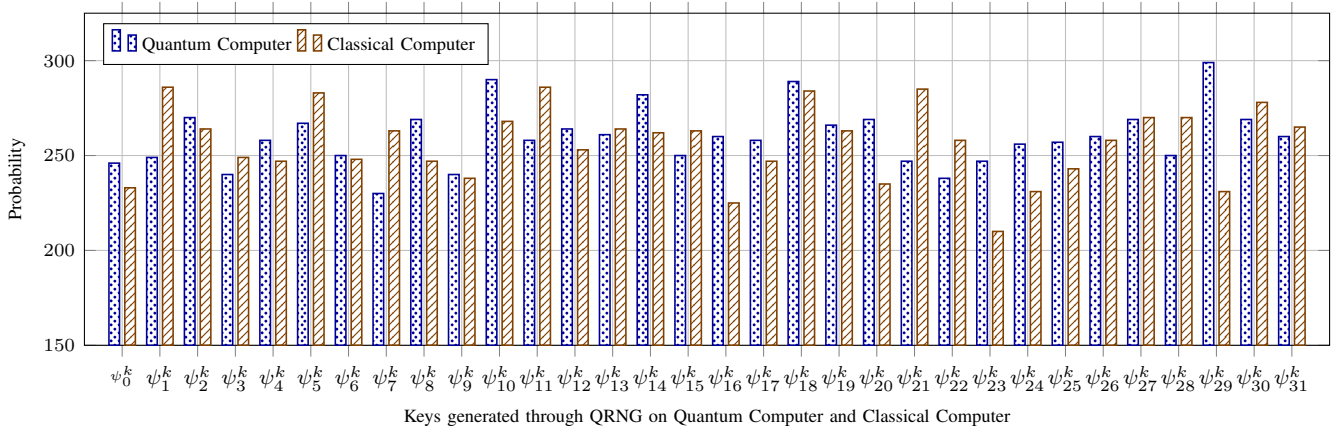
\begin{figure*}[!htbp]
\centering
\begin{tikzpicture}

\begin{axis} [filter discard warning=false, xlabel={Keys generated through QRNG on Quantum Computer and Classical Computer},ylabel={Probability}, ybar, height=6cm, width=\textwidth, style = {font=\scriptsize},
    bar width = 4pt,
    xmin=0,ymin=150,
xmax=33,ymax=325,
xtick=data,
xtick={1,2,3,4,5,6,7,8,9,10,11,12,13,14,15,16,17,18,19,20,21,22,23,24,25,26,27,28,29,30,31,32},
grid=major,
    xticklabels={\tiny $\psi_0^{k}$, \footnotesize$\psi_1^{k}$, \footnotesize $\psi_2^{k}$, \footnotesize $\psi_3^{k}$, \footnotesize $\psi_4^{k}$, \footnotesize $\psi_5^{k}$, \footnotesize $\psi_6^{k}$, \footnotesize $\psi_7^{k}$, \footnotesize $\psi_8^{k}$, \footnotesize $\psi_9^{k}$, \footnotesize $\psi_{10}^{k}$, \footnotesize $\psi_{11}^{k}$, \footnotesize $\psi_{12}^{k}$, \footnotesize $\psi_{13}^{k}$, \footnotesize $\psi_{14}^{k}$,  \footnotesize $\psi_{15}^{k}$, \footnotesize $\psi_{16}^{k}$, \footnotesize $\psi_{17}^{k}$, \footnotesize $\psi_{18}^{k}$, \footnotesize $\psi_{19}^{k}$, \footnotesize $\psi_{20}^{k}$, \footnotesize $\psi_{21}^{k}$, \footnotesize $\psi_{22}^{k}$, \footnotesize $\psi_{23}^{k}$, \footnotesize $\psi_{24}^{k}$, \footnotesize $\psi_{25}^{k}$, \footnotesize $\psi_{26}^{k}$, \footnotesize $\psi_{27}^{k}$, \footnotesize $\psi_{28}^{k}$, \footnotesize $\psi_{29}^{k}$, \footnotesize $\psi_{30}^{k}$,\footnotesize $\psi_{31}^{k}$, \footnotesize $\psi_{32}^{k}$},  
    legend style={at={(0.015,0.97)},anchor=north west,legend columns=4}, 
]
\addplot [
draw = black!40!blue,
    semithick,
    pattern = crosshatch dots,
    pattern color = black!40!blue
]   coordinates {(1, 246)(2, 249) (3, 270)(4,240) (5, 258) (6, 267) (7, 250) (8, 230) (9, 269) (10, 240)(11, 290) (12, 258) (13, 264) (14, 261) (15, 282)(16, 250) (17, 260) (18, 258) (19, 289) (20, 266) (21, 269) (22, 247) (23, 238) (24, 247) (25, 256)(26, 257) (27, 260) (28, 269) (29, 250) (30, 299) (31, 269)(32,260)};

\addplot [
draw = black!50!orange,
    semithick,
    pattern = north east lines,
    pattern color = black!50!orange
]   coordinates {(1, 233)(2, 286) (3, 264)(4,249) (5, 247) (6, 283) (7, 248) (8, 263) (9, 247) (10, 238)(11, 268) (12, 286) (13, 253) (14, 264) (15, 262)(16, 263) (17, 225) (18, 247) (19, 284) (20, 263) (21, 235) (22, 285) (23, 258) (24, 210) (25, 231)(26, 243) (27, 258) (28, 270) (29, 270) (30, 231) (31, 278)(32,265)};

\legend {Quantum Computer, Classical Computer};
\end{axis}
 
\end{tikzpicture}
\caption{Probability statistics for Key generation randomness on classical computer and classical computer.}
\label{fig:Key_Measure}
\end{figure*}

Statistical measurements of probabilities is an important metric to analyze the performance of a quantum-based algorithm, which forms a basis to access the randomness achieved in security keys along with encryption measurements. Fig. \ref{fig:Key_Measure} represents the probability of each security key that depends on number of times the QRNG is executed on quantum computer and classical computer, that turns out to be almost equivalent for each key. This equivalent comparability in their probability establishes the immunity of the encrypted data to outside attacks, while highly random keys will be more difficult to be estimated. The security keys are generated through the IBM qasm simulator and the IBM Perth quantum computer along with IBM nairobi, according to number of qubits supported.

\begin{figure}[!htbp]
\begin{tikzpicture}  
\begin{axis}  
[  
    ybar, 
    enlargelimits=0.1,
    style = {font=\scriptsize}, bar width = 7pt,
    ylabel={Probabilities}, 
    width=0.49\textwidth,height=5cm,ymax=320,
    xlabel={Encrypted States}, symbolic x coords={$\nu_0$, $\nu_1$, $\nu_2$, $\nu_3$, $\nu_4$, $\widetilde{\vert\psi\rangle}$, $\nu_5$},  
    xtick=data,  
    nodes near coords,  
    nodes near coords align={vertical},legend style={at={(0.05,0.95)},anchor=north west,legend columns=1},  
    ]  
\addplot
[
draw =green!50!blue,
     semithick,
    pattern = crosshatch,
    pattern color = green!50!blue
]  coordinates {($\nu_0$, 17) ($\nu_1$, 3) ($\nu_2$, 11) ($\nu_3$, 8) ($\nu_4$, 16) ($\widetilde{\vert\psi\rangle}$, 242) ($\nu_5$, 3)}; 
\addplot[
draw = red!50!purple,
     semithick,
    pattern = north east lines,
    pattern color = yellow!50!purple,
]  
coordinates {($\nu_0$, 0) ($\nu_1$, 0) ($\nu_2$, 0) ($\nu_3$, 0) ($\nu_4$, 0) ($\widetilde{\vert\psi\rangle}$, 300) ($\nu_5$, 0)};  
\legend{\scriptsize{Quantum Computer}, \scriptsize{Classical Computer}}    
\end{axis}  
\end{tikzpicture}
\caption{Probability statistics of measurements for encrypted states on quantum computer and classical computer}
\label{fig:QOTP_Measure}
\end{figure}
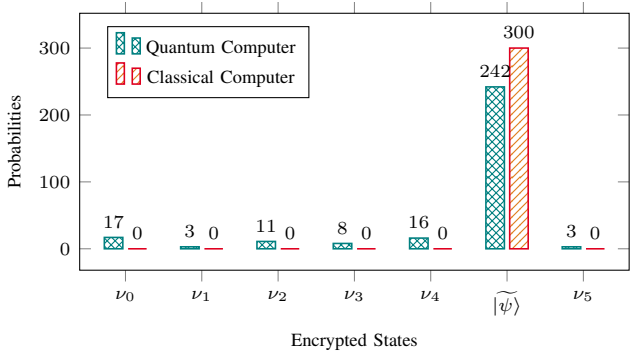

Furthermore, Fig. \ref{fig:QOTP_Measure} demonstrates the measurement performances for encryption over classical and quantum computer as well, where classical computer exhibits the precise measurements for the expected encrypted state. The quantum computer measures the expected encrypted state as maximum probable state accompanying few erroneous states ($\nu_0$-$\nu_5$) also. These error states exhibited by quantum computer is due to their fragility to external noise, decoherence and other factors impacting the qubit states. 

\begin{figure}[!htbp]
\centering
\begin{tikzpicture}
 
\begin{axis} [filter discard warning=false, xlabel={Number of instances of Consumption Data},ylabel={Computation Cost (ms)},ybar, height=5cm, width=0.49\textwidth, style = {font=\scriptsize}, 
    bar width = 6pt,
    ymin = 0,
    ymax =42,xmin=2,xmax=28,
    xtick = data,grid=major,
    ytick={10,15,20,25,30,35,40},
    legend style={at={(0.015,1.2)},anchor=north west,legend columns=3},  
]
\addplot [draw = purple!80!black,
     semithick,
    pattern = north east lines,
    pattern color = purple!80!black
]   coordinates {(5,17.79) (10,16.91)(15, 20.01) (20,24.91)(25, 26.95)};
\addplot [draw = black!40!blue,
     semithick,
    pattern = crosshatch,
    pattern color = black!40!blue
]   coordinates {(5,18.90) (10,15.24)(15, 16.30) (20,17.41)(25, 23.75)}; 
\addplot [draw = red!40!teal,
    semithick,
    pattern = north west lines,
    pattern color = red!40!teal
] coordinates {(5,28.97) (10,25.60)(15, 32.68) (20,35.15)(25, 41.05)}; 
\legend{Surveillance \cite{covid-19_surveillance_567}, TCGA \cite{glioma_grading_clinical_and_mutation_features_759}, Diabetes \cite{early_stage_diabetes_risk_prediction_dataset._529}};
\end{axis}
\end{tikzpicture}
\caption{Encryption cost with number of instances of DATA for diverse Datasets}
\label{fig:Enc_Cost}
\end{figure}
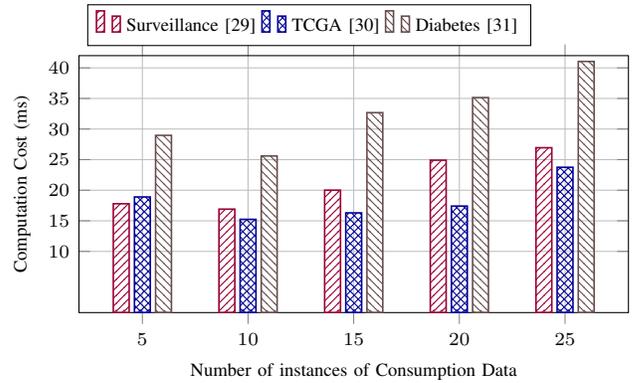

Fig. \ref{fig:Enc_Cost} provides an analytical insight to quantum encryption cost with varying number of data instances, tested on three different datasets Covid-19 surveillance \cite{covid-19_surveillance_567}, TCGA \cite{glioma_grading_clinical_and_mutation_features_759}, and Diabetes \cite{early_stage_diabetes_risk_prediction_dataset._529}. All three datasets incur a non-uniform computation cost while the Covid-19 surveillance dataset comes up with least encryption overhead and Diabetes dataset with the maximum encryption overhead. 
\subsubsection{Prediction Result}\label{subsecpred}

A coherent insight to quantum neural network driven prediction frameworks' loss parameter; over different number of epochs, in two different scenarios with two and four qubits, for dataset size 2k and 10k, respectively is depicted in Fig. \ref{fig_loss}. It is visible from figure that, framework is performing better with increase number of data instance as it can learn, significantly from a more informed data. Moreover, prediction unit performance is appearing better by reducing the loss value, with increased number of qubits due to deep computation with increase number of qubits. 

\begin{figure*}[!htbp]
\centering
\begin{subfigure}
\centering
\begin{tikzpicture}[node distance = 1.0cm, auto,scale=.950, transform shape]
\begin{axis}[filter discard warning=false,grid=major,grid style={dashed}, width=0.99\columnwidth, height=5.0cm,
xmin=1,xmax=100, ylabel={Loss},xlabel={Epochs},legend style={at={(0.35,0.9)},anchor=west},legend columns=4, style = {font=\scriptsize}]
\addplot [draw = black!50!orange, style={dashed}, thick] table[x expr= 1*\coordindex,y=L22]{values2k.dat};
\addplot [draw = black!50!teal, thick] table[x expr= 1*\coordindex,y=L24]{values2k.dat};
\legend{$2Q$,$4Q$}
\end{axis}
\end{tikzpicture}
\end{subfigure}
\hfill
\begin{subfigure}
\centering
\begin{tikzpicture}[node distance = 1.0cm,auto,scale=.950, transform shape]
\begin{axis}[filter discard warning=false,grid=major,grid style={dashed}, width=0.99\columnwidth, height=5.0cm,
xmin=1,xmax=100, ylabel={Loss},xlabel={Epochs},legend style={at={(0.35,0.9)},anchor=west},legend columns=4, style = {font=\scriptsize}]
\addplot [draw = black!50!orange, style={dashed}, thick] table[x expr= 1*\coordindex,y=L2]{values10k.dat};
\addplot [draw = black!50!teal, thick] table[x expr= 1*\coordindex,y=L4]{values10k.dat};
\legend{$2Q$,$4Q$}
\end{axis}
\end{tikzpicture}
\end{subfigure}
\caption{Prediction loss values with Epochs = 100. (a) 2k (b) 10k}
\label{fig_loss}
\end{figure*}
\subsection{Comparison}\label{subseccomp}
IQ-HDM is compared with existing state-of-the-art works like \textit{Machine Learning and Probabilistic Analysis Based Model} (MLPAM) \cite{gupta2020mlpam}, \textit{Intelligent Security Performance Prediction for IoT-Enabled Healthcare Networks Using an Improved CNN} (IoT-HSM) \cite{IOTHSM9439861}, \textit{Quantum Machine Learning driven Malicious User Prediction Model} (QM-MUP) \cite{QM-MUP9865138}, \textit{Malicious Agent Identification-based Data Security Model for cloud environments} (MAIDS) \cite{MAIDS}, and \textit{Federated learning driven Malicious User Prediction Model for secure data distribution in cloud environments} (FedMUP) \cite{FedMUPGUPTA2024111519}. Brief details regarding these state-of-the-art works are already discussed in Section \ref{subsecrel}.

\par
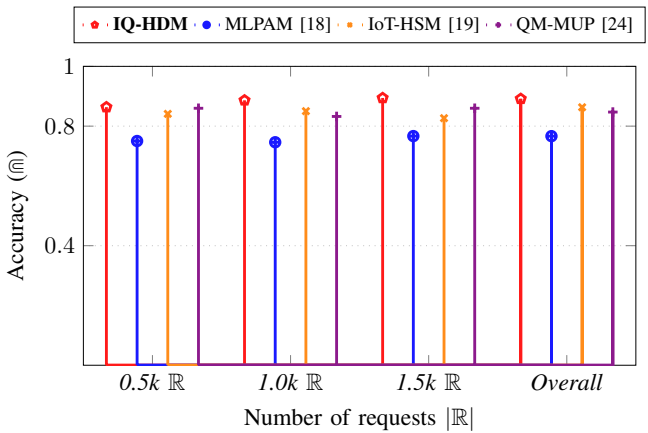
\begin{figure}[!htbp]
\begin{tikzpicture}[node distance = 1cm,auto,scale=.90, transform shape]
\pgfplotsset{every axis y label/.append style={rotate=180,yshift=10.5cm}}
\begin{axis}[
      axis on top=false,
        height=6.0cm,
        width=1.1\linewidth,
      xmin=9, xmax=81,
      ymin=0, ymax=1,
      ytick={0.4,0.8,1},
      xtick={18,36,54,72},
      xticklabels={\textit{0.5k $\mathbb{R}$}, \textit{1.0k $\mathbb{R}$}, \textit{1.5k $\mathbb{R}$}, \textit{Overall}},
      xlabel={Number of requests $|\mathbb{R}|$},
        ycomb,
        ylabel near ticks, yticklabel pos=left,
      ylabel={Accuracy ($\doublecap$)},
       legend image post style={scale=0.6}, 
       legend style={at={(0.5,1.20)},
      anchor=north,legend columns=4},
      ymajorgrids=true,
      grid style=dotted,
          ]
\addplot+[mark=pentagon, mark options={fill=red},fill=red, draw=red!90,  very thick] 
coordinates
{(12,.8635) (30,.8864) (48,.8942) (66,.8914) }
\closedcycle;
\addlegendentry{\footnotesize \textbf{IQ-HDM}}

\addplot+[mark=oplus,mark options={fill=blue},fill=blue,draw=blue!90, very thick]
coordinates
 {(16,.7502) (34,.7461) (52,.7667) (70,.7665) }
\closedcycle;
\addlegendentry{\footnotesize MLPAM \cite{gupta2020mlpam}}

\addplot+[mark=x,mark options={fill=pink},fill=pink,draw=orange!90, very thick] 
coordinates
 {(20,.8411) (38,.8502) (56,.8267) (74,.8632) }
\closedcycle;
\addlegendentry{\footnotesize IoT-HSM \cite{IOTHSM9439861}}

\addplot+[mark=+,mark options={fill=white},fill=red,draw=violet!90, very thick] 
coordinates
 {(24,.8598) (42,.8324) (60,.8598) (78,.8471) }
\closedcycle;
\addlegendentry{\footnotesize QM-MUP \cite{QM-MUP9865138}}
\end{axis}
\end{tikzpicture}
\caption{Comparative analysis of accuracy with state-of-the-art approaches.} \label{figcomp}
\end{figure}

 Fig. \ref{figcomp} depict the comparison of the proposed framework's performance parameters; accuracy and data breach coverage with other state-of-the-art works by considering different data access request scenarios for \{0.5k, 1.0k, 1.5k, Overall\} requests, respectively. The accuracy of the proposed approach is having an edge over all the compared approaches for different request scenarios. High values of $\doublecap$ show the improved performance in range between 3.13\% to 16.13\%. Hence, this is evident that the IQ-HDM is outperforming considered existing approaches and its performance is remarkably elevated due to secure storage, communication, and prediction strategies in the proposed framework.


Table \ref{tab4} entails an absolute deviation among proposed IQ-HDM and existing models MLPAM, IoT-HSM, QM-MUP, MAIDS, and FedMUP by correlating different security features. It exhibits that IQ-HDM is the only framework to assume all entities not fully trusted. This means any of the entity can have mal-intentions and responsible for the data breaches. Also, it facilitates potent security and privacy features such as data storage, data communication and malicious prediction altogether delivers comprehensively secure communication and visionary breach prediction. 
\begin{table}[!htbp]
\caption{Feature Analysis: Proposed Vs Existing models}
\label{tab4}
\setlength{\tabcolsep}{3pt}
\begin{tabular}{p{55pt}|p{20pt}p{20pt}p{20pt}p{20pt}p{20pt}|p{55pt}}
\hline
\textbf{Models}& \textbf{$\mathbb{O}$} & \textbf{$\mathbb{SC}$} & \textbf{$\mathbb{SDS}$} & \textbf{$\mathbb{SDD}$} & \textbf{$\doublecap$} & \textbf{Complexity} \\
\hline \hline
MLPAM \cite{gupta2020mlpam} & $\star$ & $\times$ & $\surd$ & $\surd$ & 76.65 & $\mathcal{O}(|\sum_{j=1}^{m}d_{j}|)$  \\
IoT-HSM \cite{IOTHSM9439861} & $\star\star$ & $\surd$ & $\surd$ & $\times$ & 86.32 & $\mathcal{O}(|\sum_{j=1}^{m}d_{j}|)$   \\ 
QM-MUP \cite{QM-MUP9865138} & $\times$ & $\surd$ & $\surd$ & $\surd$ & 84.71 & $\mathcal{O}$($tL\mathcal{N}$$N^{\ast})$ \\
MAIDS\cite{MAIDS} & $\times$ & $\surd$ & $\times$ & $\surd$ & 86.75 & ${O}$($tmxyN)$ \\
FedMUP \cite{FedMUPGUPTA2024111519} & $\times$ & $\surd$ & $\times$ & $\surd$ & 87.24 & ${O}(ntL\xi{N}^{\ast})$ \\
\textbf{IQ-HDM} & $\star\star$ & $\surd$ & $\surd$ & $\surd$ & 89.02 & 
		$\mathcal{O}(nQE\Tilde{Q}N^{\ast})$ \\
\hline
\multicolumn{7}{p{240pt}}{$\star$: Single; $\star\star$: Multiple, $\mathbb{O}$: IoT Devices, $\mathbb{SC}$: Secure Communication, $\mathbb{SDS}$: Secure Data Storage, $\mathbb{SDD}$: Secure Data Distribution, $\doublecap$: Accuracy.}
\end{tabular}
\label{tab2}
\end{table}
The proposed quantum driven healthcare data management framework outperforms the state-of-the-art data security methods, and it is suitable to enhance the performance of breach prediction in a distributed environment. The reason for this enhanced performance of predicted values is the learning of useful information by using quantum values from input data samples.

After comprehensive evaluation of proposed work the enhancements in terms of quantum network security observed are that the proposed framework provisions an unconditional security through QOTPE and OPHDC by utilizing the quantum mechanical principles making it highly immune to classical and quantum computer attacks as well, unlike classical security mechanisms relying on computational hardness problems. Moreover, the sensitive data after being encrypted through QOTPE turns out to be maximally mixed states, that are completely unidentifiable by any adversary. Any measurements or alterations made by adversary can be detected trivially. Consequently, IQ-HDM stands ahead in all respects for supporting secure healthcare data access and cloud communication for overall data management. 
\subsection{Discussion}\label{secdiscuss}
The results of this study showcase the effectiveness of the proposed approach. Extant literature reflects that none of the existing approaches alone is sufficient to impart healthcare data management. Accordingly, the study proposed a comprehensive framework to ensure all-round data management by utilizing the one of finest tools, quantum computing. The rationality to be motivated for the deployment of a quantum-oriented data security approach lies in the fact that quantum approaches are far superior to classical computing as quantum gate permits an infinite number of qubit states and rotational outcome in a 360$^{\circ}$ view for deep analysis to predict the data breach, efficiently. The quantum one-time padding encrypts the data before storage on the cloud and then the quantum malicious entity prediction unit analyses the user intention before allocating data. Hence, the study first fortified the crucial data and then checked for user intention. The significant impact of the study is in the successful implementation to deliver comprehensive healthcare data management.
\section{Conclusion and Future Work}\label{seccon}
Quantum oriented comprehensive data management framework is proposed to provide secure data storage, implementing data privacy and security policy with expanded digitalization of healthcare data. The IQ-HDM framework utilizes the QOTPE unit to enhance the quality of data-sharing needs and the QPHDC unit to strengthen data communication for proactive estimation of the malicious entity. In this way, the framework furnishes a more nuanced and contextually relevant approach in the context of healthcare data security. Also, extensive experimental work has been conducted to demonstrate the effectiveness of the proposed methodology in real-world scenarios, contributing insights into the application of these criteria in the healthcare domain. This ensures that the proposed work goes beyond mere theoretical alignment with established practices, offering a substantively practical contribution to the field. 

In the future, the IQ-HDM framework can be extended to develop an advanced, robust, and effective mechanism to enhance its capability of detecting the malicious entity, in case crucial data got disclosed intentionally or non-intentionally. Additionally, quantum-based transfer learning can be utilized to improve the proposed framework by making it capable of countering unknown types of cyber attacks. 


\bibliographystyle{IEEEtran}
\bibliography{reference}


\begin{IEEEbiography}[{\includegraphics[width=1in,height=1.25in,clip,keepaspectratio]{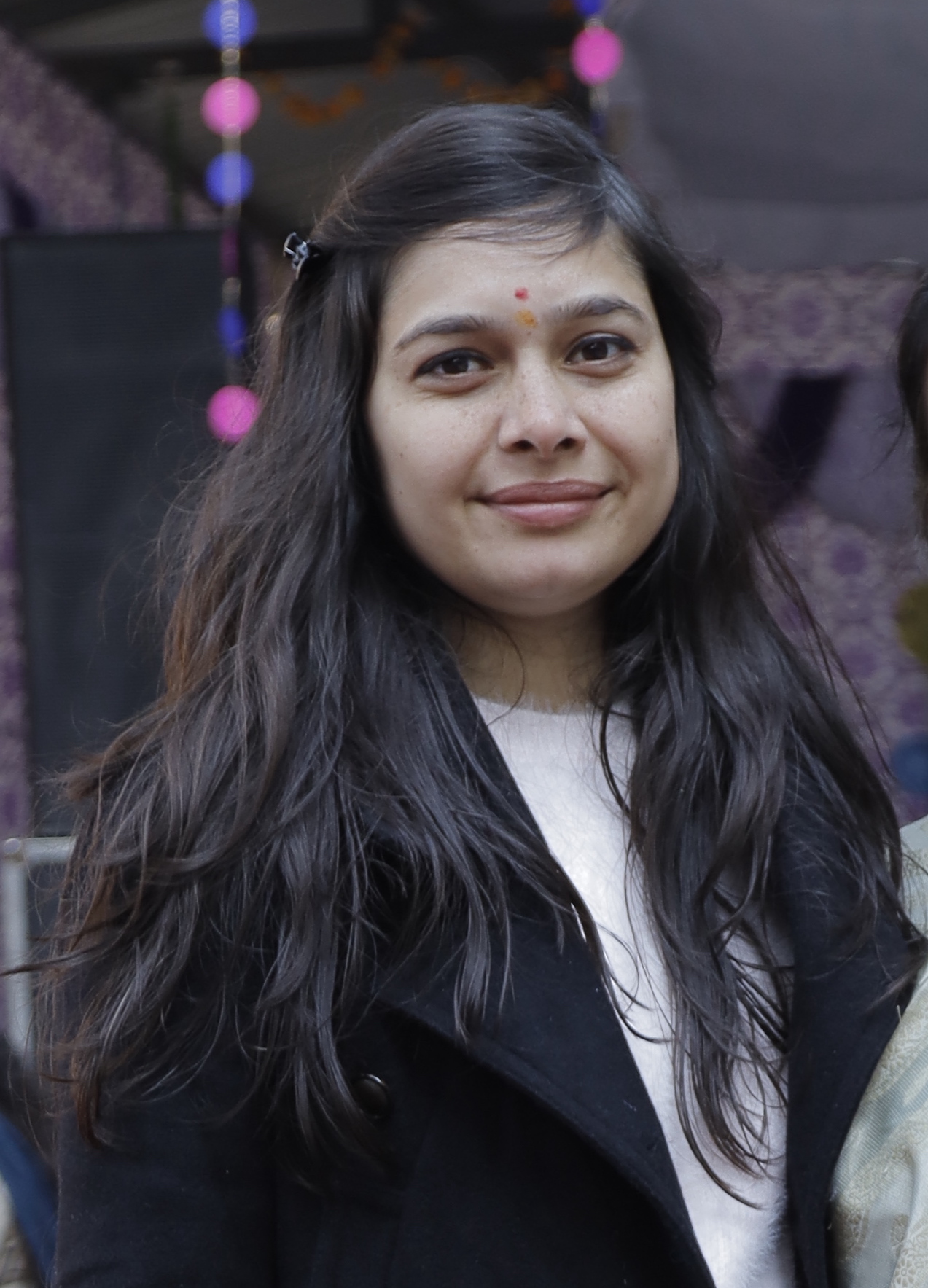}}]{Kishu Gupta} (Member, IEEE) received Ph.D. degree form India in 2023. She is currently a Post-Doctoral Research Fellow with the Cloud Computing Research Center, Department of Computer Science and Engineering, National Sun Yat-sen University (NSYSU), Kaohsiung, Taiwan. Her research interests include data security and privacy, cloud computing, traffic management, federated learning, machine learning, neural networks, and quantum machine learning. She has research findings published with top-notch venues, such as IEEE TRANSACTIONS ON CONSUMER ELECTRONICS, Applied Soft Computing, and Cluster Computing. She was a recipient of the Gold Medal for securing first rank in overall university during the M.Sc. degree (computer science). She received the prestigious INSPIRE Fellowship sponsored by the Department of Science and Technology (DST) under the Ministry of Science and Technology, Government of India, for her Ph.D. degree.
\end{IEEEbiography}
\vspace{11pt}
\begin{IEEEbiography}[{\includegraphics[width=1in,height=1.25in,clip,keepaspectratio]{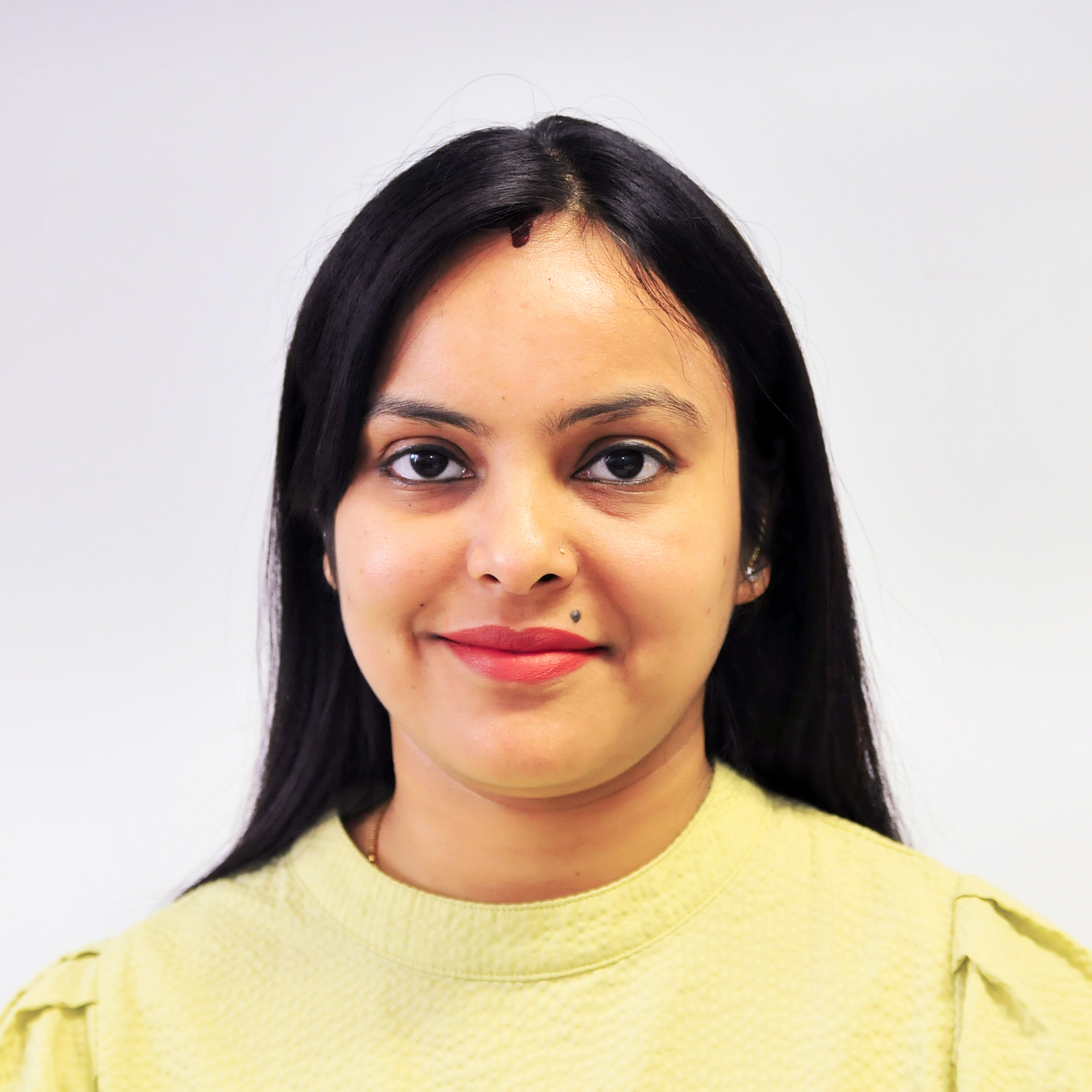}}]{Deepika Saxena} (Member, IEEE) received the Ph.D. degree in computer science from the National Institute of Technology, Kurukshetra, India. She was a Post-Doctoral Research Fellow with the Department of Computer Science, Goethe University, Frankfurt, Germany. She is currently an Associate Professor with the Division of Information Systems, The University of Aizu, Japan. Also, she is an Online Lecturer with the University of Economics and Human Sciences, Warsaw, Poland, Europe. Her research interests include neural networks, evolutionary algorithms, resource management and security in cloud computing,	internet traffic management, quantum machine learning, data lakes, and	dynamic caching management. She was a recipient of the prestigious IEEE TCSC 2023 Outstanding Ph.D. Dissertation Award and the EUROSIM 2023	Best Ph.D. Thesis Award. She received the 2022 Best Paper Award for her	research article published in IEEE TRANSACTIONS ON CLOUD COMPUTING JOURNAL. Also, she is the recipient of the prestigious Japan Society for the Promotion of Science (JSPS) KAKENHI Early Career Young Scientist Research Grant FY2024.
\end{IEEEbiography}
\vspace{11pt}
\begin{IEEEbiography}[{\includegraphics[width=1in,height=1.25in,clip,keepaspectratio]{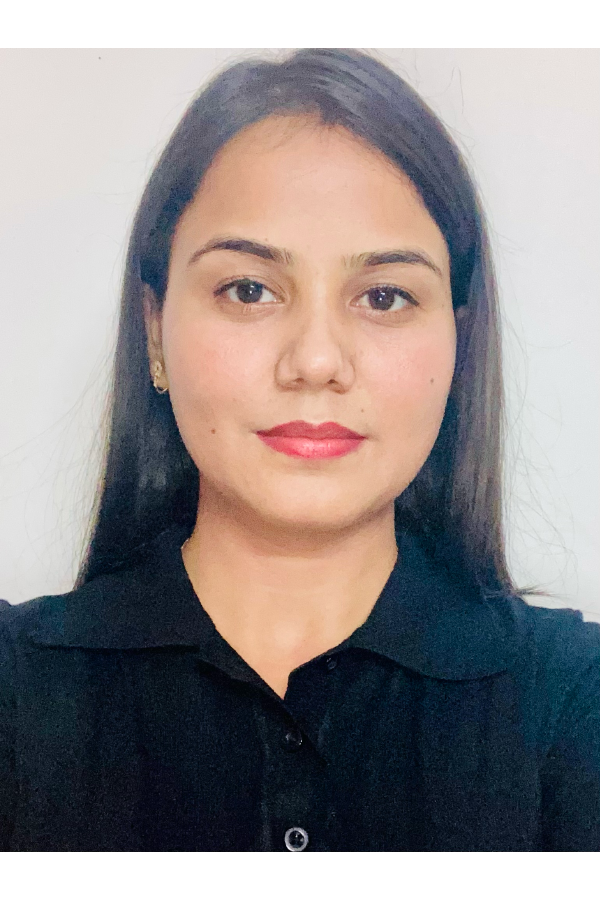}}]{Pooja Rani} received the M.Sc. degree from the Department of Computer Science and Applications, Kurukshetra University, Kurukshetra, Haryana, India. She is currently pursuing the Ph.D. degree with the Department of Computer Applications, National Institute of Technology, Kurukshetra. Her research interests include data security and demand-response management in smart grids, predictive analytics, quantum computing, and cloud computing.
\end{IEEEbiography}
\vspace{11pt}
\begin{IEEEbiography}[{\includegraphics[width=1in,height=1.25in,clip,keepaspectratio]{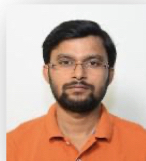}}]{Jitendra Kumar} (Senior member, IEEE) received the Ph.D. degree in machine learning and cloud computing from the National Institute of Technology Kurukshetra, India, in 2019. He is currently an Assistant Professor with the Department of Mathematics, Bioinformatics, and Computer Applications, Maulana Azad National Institute of Technology Bhopal, India. His research interests include cloud computing, computational intelligence, time series forecasting, and optimization.
\end{IEEEbiography}
\vspace{11pt}
\begin{IEEEbiography}[{\includegraphics[width=1in,height=1.25in,clip,keepaspectratio]{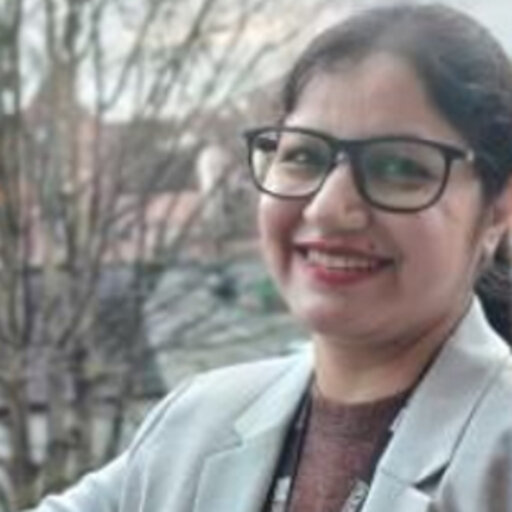}}]{Aaisha Makkar} (Member, IEEE)received the Ph.D. degree from the Thapar Institute of Technology, Patiala, India. She was a Post-Doctoral (Research) Professor with the Department of Computer Science, Seoul National University of Science and Technology, South Korea. She is currently a Lecturer of computer science with the College of Science and Engineering, University of Derby, U.K. Her current research interests include cloud computing, machine learning, data analytics, and parallel processing.
\end{IEEEbiography}
\vspace{11pt}
\begin{IEEEbiography}[{\includegraphics[width=1in,height=1.25in,clip,keepaspectratio]{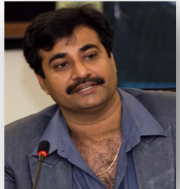}}]{Ashutosh Kumar Singh} (Senior Member, IEEE) received the Ph.D. degree in electronics engineering from Indian Institute of Technology (BHU) Varanasi, India. He was a Post-Doctoral Researcher with the Department of Computer Science, University of Bristol, U.K. He is currently a Professor and the Director of Indian Institute of Information Technology Bhopal, India. Also, he is an Adjunct Professor with the University of Economics and Human Sciences, Warsaw, Poland. He has research and teaching experience in various universities in India, the U.K., and Malaysia. He has published more than 400 research papers in different journals and conferences of high repute. Some of his research findings are published in top cited journals, such as IEEE TRANSACTIONS ON SERVICES COMPUTING, IEEE TRANSACTIONS ON COMPUTERS, IEEE TRANSACTIONS ON SYSTEMS, MAN, AND CYBERNETICS, IEEE TRANSACTIONS ON PARALLEL AND DISTRIBUTED SYSTEMS, IEEE TRANSACTIONS ON INDUSTRIAL INFORMATICS, IEEE TRANSACTIONS ON CLOUD COMPUTING, IEEE COMMUNICATIONS LETTERS, IEEE NETWORKING LETTERS, IEEE DESIGN AND TEST, IEEE SYSTEMS JOURNAL, IEEE WIRELESS COMMUNICATION LETTERS, IEEE TRANSACTIONS ON NETWORK AND SERVICE MANAGEMENT, IEEE TRANSACTIONS ON GREEN COMMUNICATIONS AND NETWORKING, IET Electronics Letters, FGCS, Neurocomputing, Information Sciences, and Information Processing Letters. His research interests include the design and testing of digital circuits, data science, cloud computing, machine learning, and security. His research paper, published in IEEE TRANSACTIONS ON CLOUD COMPUTING JOURNAL was honored with the 2022 Best Paper Award by the IEEE Computer Society Publications Board. 
\end{IEEEbiography}
\vspace{11pt}
\begin{IEEEbiography}[{\includegraphics[width=1in,height=1.25in,clip,keepaspectratio]{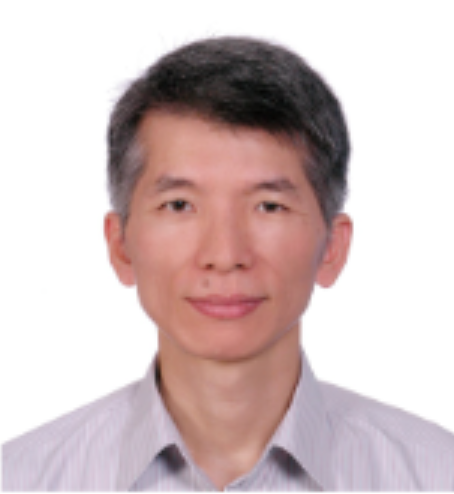}}]{Chung-Nan Lee} (Member, IEEE)received the B.S. and M.S. degrees in electrical engineering from National Cheng Kung University, Tainan, Taiwan, in 1980 and 1982, respectively, and the Ph.D. degree in electrical engineering from the University of Washington, Seattle, WA, USA, in 1992. Since 1992, he has been with National Sun Yat-sen University, Kaohsiung, Taiwan, where he was the Chairperson of the Department of Computer Science and Engineering from 1999 to 2001. Currently, he is a Distinguished Professor and the Director of the Cloud Computing Research Center. His research interests include multimedia over wireless networks, cloud computing, and the IoT. He was the President of the Taiwan Association of Cloud Computing from 2015 to 2017 and the VP for TA of Asia–Pacific Signal and Information Processing Association from 2019 to 2020. In 2016, he received the Outstanding Engineering Professor Award from Chinese Institute of Engineers, Taiwan. 
\end{IEEEbiography}

\end{document}